\documentclass[aps,prl,reprint,groupedaddress,nofootinbib,floatfix]{revtex4-1}
\usepackage{amssymb,amsmath,amsthm}
\usepackage{xcolor}
\usepackage{hyperref}
\usepackage{slashed}
\usepackage{bm}

\usepackage[utf8]{inputenc}

\newtheorem{conj}{Conjecture}

\newcommand{\df}{:=}
\newcommand{\iu}{{\rm i}}

\begin{document}

\title{Emergence and the Swampland Conjectures}

\author{Ben Heidenreich}
\email{bheidenreich@perimeterinstitute.ca}
\affiliation{Perimeter Institute for Theoretical Physics, Waterloo, Ontario, Canada N2L 2Y5}

\author{Matthew Reece}
\email{mreece@physics.harvard.edu}
\affiliation{Department of Physics, Harvard University, Cambridge, MA, 02138}

\author{Tom Rudelius}
\email{rudelius@ias.edu}
\affiliation{School of Natural Sciences, Institute for Advanced Study, Princeton, NJ 08540, USA}

\date{\today}

\begin{abstract}
The Ooguri-Vafa Swampland Conjectures claim that in any consistent theory of quantum gravity, when venturing to large distances in scalar field space, a tower of particles will become light at a rate that is exponential in the field space distance. We provide a novel viewpoint on this claim: if we assume that a tower of states becomes light near a particular point in field space, and we further demand that loop corrections drive both gravity and the scalar to strong coupling at a common energy scale, then the requirement that the particles become light exponentially fast in the field-space distance in Planck units follows automatically. Furthermore, the same assumption of a common strong-coupling scale for scalar fields and gravitons  implies that when a scalar field evolves over a super-Planckian distance, the average particle mass changes by an amount of order the cutoff energy. This supports earlier suggestions that significantly super-Planckian excursions in field space cannot be described within a single effective field theory. We comment on the relationship of our results to the Weak Gravity Conjecture.
\end{abstract}

% insert suggested PACS numbers in braces on next line
% insert suggested keywords - APS authors don't need to do this
%\keywords{}

\maketitle

\section{Introduction}

A question of fundamental interest in the study of quantum gravity is: which low-energy gravitational effective field theories (EFTs) admit an ultraviolet (UV) completion? Said differently, what are the low-energy predictions of quantum gravity?

String theory provides a very large class of quantum gravities, with a wide variety of possible low-energy EFTs (see, e.g., \cite{Douglas:2004zg, Taylor:2015xtz}). But despite this enormous ``landscape'' of vacua, there is growing evidence that they all share identifiable common features, distinguishing them from ``swampland''~\cite{Vafa:2005ui}, a large class of seemingly-consistent gravitational EFTs with no quantum gravity UV completion. Thus, our original question becomes: how do we distinguish the swampland from the landscape?

There are several candidate criteria for fencing off parts of the swampland. In this paper, we focus on the ``Swampland Conjectures'' of~\cite{Vafa:2005ui,Ooguri:2006in} concerning the moduli space of a quantum gravity. We will apply ideas about effective field theory and the infrared emergence of weak coupling recently developed in~\cite{Heidenreich:2017sim} (see also~\cite{Harlow:2015lma}), where they were used to understand another candidate swampland criterion, the weak gravity conjecture (WGC)~\cite{ArkaniHamed:2006dz}. Other recent works on the Swampland Conjectures include~\cite{Baume:2016psm, Klaewer:2016kiy, Blumenhagen:2017cxt, Palti:2017elp, Hebecker:2017lxm}.

Controlled string compactifications typically have many ``moduli'': light scalars with gravitational-strength couplings. This is related to the fact that string theory has no continuous parameters, so any freely adjustable coupling must be controlled by the vacuum expectation value (vev) of a scalar field. In the simplest supersymmetric examples the moduli are exactly massless and parameterize a continuous moduli space of vacua. More realistic examples require a moduli potential with isolated minima---the vacua of the theory---but the only reliable way to generate vacua in the weak-coupling regime is through non-perturbative effects. Since these are exponentially small at weak coupling, the moduli masses are likewise exponentially suppressed relative to, e.g., the Kaluza-Klein (KK) scale or the string scale.

``Conjecture 0'' of~\cite{Ooguri:2006in} generalizes these observations to any quantum gravity:
\setcounter{conj}{-1} %ensure conjectures are numbered 0,1,2
\begin{conj} % #0
Every continuous parameter in a quantum gravity is controlled by the vev of a scalar field.
\end{conj}
\noindent Thus, we expect moduli to be ubiquitous in the landscape, and it is natural to investigate the properties of the moduli space of a quantum gravity. In particular, we focus on two related conjectures from~\cite{Ooguri:2006in}:
\begin{conj} % #1
The moduli space has infinite diameter (despite finite volume~\cite{Vafa:2005ui,Douglas:2005hq}).
\end{conj}
\begin{conj} % #2
At large distances in moduli space, an infinite tower of resonances becomes light exponentially quickly with increasing distance.
\end{conj}
\noindent Here distances in moduli space are defined by geodesic distances with respect to the moduli space metric $g_{i j}(\phi)$:
\begin{equation}
\mathcal{L}_{\rm kin}^{(\mathrm{mod})} = \frac{1}{2} g_{i j}(\phi) \partial \phi^i \cdot \partial \phi^j.
\end{equation}
Implicit in these conjectures is the notion that parametrically distant points in moduli space correspond to weak-coupling limits.

In this paper, we explore the connection between weak coupling, a tower of light resonances, and large distances in moduli space. Central to our arguments is the assumption, developed in~\cite{Heidenreich:2017sim}, that weak coupling in quantum gravities is a long-distance phenomenon, with all physics becoming strongly coupled in the ultraviolet at a common ``quantum gravity scale.'' This occurs in many concrete quantum gravities, such as large volume compactifications of M-theory. While there are possible counterexamples as well, see~\cite{Heidenreich:2017sim}, a suitable generalization of this notion may address these, and it is worthwhile to understand the consequences regardless.

This assumption clarifies the connection between light resonances and weak coupling. For example, light charged resonances screen gauge forces, leading to weak gauge couplings in the infrared: this is essentially the mechanism of~\cite{Heidenreich:2017sim}.

We now explain the connection between light resonances and parametrically large distances in moduli space, applying the same assumption. Applying similar reasoning, we further consider the effect of moving a large but bounded distance in moduli space, with particular attention to the case of an axion with a transplanckian decay constant.\footnote{While this paper was in preparation (following a strategy sketched in the conclusions of \cite{Heidenreich:2017sim}), we became aware of independent work~\cite{PaltiValenzuela} with some overlapping results.}

\section{Light resonances and large distances}

Motivated by Conjecture 2, we consider a trajectory in moduli space approaching a singular point where an infinite tower of resonances becomes massless. This trajectory is parameterized by a scalar field $\phi$; we take the singular point to be $\phi = 0$ without loss of generality. We assume that the spectrum is dominated by an infinite tower of particles that become \emph{uniformly} light as $\phi\to0$, so
\begin{equation}
m_n \approx \mu_n \phi + {\cal O}(\phi^2)
     \label{eq:tower}
\end{equation}
after an appropriate redefinition of $\phi$.
 In making this ansatz, we are no longer free to assume that $\phi$ has a canonical kinetic term.

The ansatz~\eqref{eq:tower} is quite general. For instance, suppose there are multiple  towers with masses trending to zero at different rates. Then, by moving far enough in the moduli space, we can focus on the tower that becomes light fastest and parameterize the modulus $\phi$ so that this tower becomes light in linear fashion. Thus, aside from being justified by specific examples, we believe that \eqref{eq:tower} captures the most general intended meaning of ``tower'' in Conjecture 2.

For concreteness we take the particles in the tower to be Dirac fermions, though similar results apply for other spins. The Lagrangian is then
\begin{equation}
{\cal L} = \frac{1}{2}K(\phi)(\partial \phi)^2 - V(\phi) + \sum_n \left[\overline{\psi}_n \left(\iu \slashed{\partial} - \mu_n \phi\right)\psi_n\right].   
\label{eq:Lfermions}
\end{equation}
Fermion loops will correct the $\phi$ propagator as well as $S$-matrix elements for scattering of $\phi$ particles. At some energy scale $\Lambda$ these loop corrections  become as large as the tree-level contribution and the EFT breaks down. We follow the approach of \cite{Heidenreich:2017sim} to ascertain the scale $\Lambda$, with the new ingredient that $\Lambda(\phi)$ depends on the value of the modulus.\footnote{We work in Einstein frame, holding the $D$-dimensional Newton's constant fixed as $\phi$ varies. Alternatively, one could hold the cutoff $\Lambda$ fixed and view the strength of gravity as varying across the moduli space, which may be more natural when $\phi \to 0$ corresponds to a decompactification limit.}

We now compute the loop correction to the $\phi$ propagator from the tower of fermions, perturbing around a fixed expectation value $\langle\phi\rangle = \phi_0$.\footnote{For general $V(\phi)$ and general $\phi_0$ there is a tadpole and the field will begin to roll. Implicit in the notion of a modulus is that the potential is relatively flat, hence there is a finite time scale over which our computation of scattering with fixed $\langle\phi\rangle = \phi_0$ makes sense. We comment further on the flatness of the potential below.} The canonically normalized fluctuation $\overline{\phi} \df \sqrt{K(\phi_0)}(\phi - \phi_0)$ has propagator
\begin{equation}
 \langle \widetilde{\overline{\phi}} (p)  \widetilde{\overline{\phi}} (- p) \rangle \sim \frac{1}{p^2 - m_\phi^2 + \iu
  \varepsilon}  \frac{1}{1 + \Pi (p^2)} .
\end{equation}
By adding appropriate counterterms, we can choose the
renormalization condition $\Pi (0) = 0$ (up to possible infrared divergences if there is a massless particle coupling to $\phi$).

Parametrically, the large-$p$ behavior of the contribution to the one-loop integral from fermion $n$ is
\begin{equation}
  \left| \Pi_n (p^2 \gg m_n^2) \right| \sim \frac{\mu_n^2}{K(\phi_0)} p^{D - 4} \sim \frac{1}{K(\phi_0)} \biggl(\frac{\partial m_n}{\partial \phi}\biggr)^2 p^{D - 4}.
\end{equation}
To assess the strong coupling scale, we define a function $\lambda_\phi(p)$ that captures the parametric contribution to $\Pi(p^2)$ from the sum over {\em only} those particles with mass less than $p$. We have
\begin{equation}
  \lambda_{\phi} (p) \df \frac{p^{D - 4}}{K (\phi_0)} \sum_{n \, | \, m_n(\phi_0) < p}
  \left( \frac{\partial m_n}{\partial \phi} \right)^2.
  \label{eq:lambdadef}
\end{equation}
From this expression one can read off the strong coupling scale $\Lambda(\phi_0)$ as the value of $p$ where $\lambda_\phi(p) \sim 1$. The result clearly depends on the unknown function $K(\phi_0)$.

\section{Strong coupling and gravity}

The loop corrections from the tower of light particles also affect the graviton propagator, leading to strong coupling \cite{Dvali:2007hz,Dvali:2007wp}. These loop corrections are parametrically controlled by
\begin{equation}
 \lambda_{\rm grav}(p) \df G_N \,p^{D-2} N(p),
\end{equation}
where $N(p) =  \sum_{n \, | \, m_n(\phi_0) < p} 1$ is the number of weakly coupled particles with mass below $p$.

Consider the ``species bound'' scale $\Lambda_{\rm QG}(\phi_0)$ at which $\lambda_{\rm grav}(p) \sim 1$:
\begin{equation}
\Lambda^{D-2}_{\rm QG} = \frac{1}{G_N N(\Lambda_{\rm QG})}.
\label{eq:speciesbound}
\end{equation}
In terms of this scale we have
\begin{equation}
\lambda_\phi(\Lambda_{\rm QG}) = \frac{1}{\Lambda_{\rm QG}^2 K(\phi_0) G_N} \left<\mu^2\right>_{\Lambda_{\rm QG}},
\label{eq:lambdaphi}
\end{equation}
where 
\begin{align}
\left<\mu^2\right>_\Lambda &\df \frac{1}{N(\Lambda)} \sum_{n\, |\, m_n(\phi_0) < \Lambda} \left(\frac{\partial m_n}{\partial \phi}\right)^2 \nonumber \\
&\approx \frac{1}{N(\Lambda) \phi^2}  \sum_{n\, |\, m_n(\phi_0) < \Lambda} m_n^2 \sim \frac{\Lambda^2}{\phi^2}.
\end{align}
In the last line we have made a mild but crucial assumption that most of the particles in the tower lie near the cutoff, as in any tower with an increasing density of states $d N/d p$, or with a power-law (increasing or decreasing) density of states.

Given this assumption, if we demand that the modulus becomes strongly coupled at the ``quantum gravity scale'' $\Lambda_{\rm QG}$ in \eqref{eq:speciesbound}, we constrain the form of the kinetic term:
\begin{equation}
\lambda_\phi(\Lambda_{\rm QG}) \sim 1 \Rightarrow K(\phi_0) \sim \frac{1}{G_N \phi_0^2}.
\end{equation}
That is, for a wide variety of spectra for the tower of states that become light as $\phi \to 0$, the condition that the modulus becomes strongly coupled at the quantum gravity scale fixes the kinetic term of the modulus to be
\begin{equation}
{\cal L}_{\rm kin} \sim M_{\rm Pl}^{D-2} \frac{1}{\phi^2} (\partial \phi)^2. 
   \label{eq:finalkinetic}
\end{equation}
In particular, distances in field space grow logarithmically with the value of $\phi$. Equivalently, the particles in the tower become exponentially light in the field-space distance, precisely as required by Conjecture 2. Slightly different calculations give the same results for a tower of scalars;\footnote{There are some subtleties related to counterterm contributions in $D = 4$, which however do not change the result.} importantly, unlike mass corrections, these kinetic corrections do not cancel between scalars and fermions in supersymmetric theories.

Under similar assumptions about the density of states within the tower, we obtain the energy-dependent statement $\sum_{n\, |\, m_n(\phi_0) < p} m_n^2 \sim N(p)\, p^2$ for any $p \gg p_0$, with $p_0$ some characteristic scale (e.g., the lowest mass threshold in the tower).
Applying \eqref{eq:finalkinetic}, we find
\begin{equation}
\lambda_\phi(p) \sim G_N N(p) p^{D-2} \sim \lambda_{\rm grav}(p), \quad p\gg p_0. 
\end{equation}
This can be interpreted as a type of ``unification'' of the strengths of loop effects for gravity and for the modulus at energies above $p_0$, similar to the unification of gauge and gravity loops discussed in \cite{Heidenreich:2017sim}. 

 Note that \eqref{eq:finalkinetic} implies that the modulus has gravitational-strength couplings to the particles in the tower; this is a well-known characteristic of moduli, here a natural consequence of our assumptions.

\section{Transplanckian distances and EFT}

Conjecture 2 implies that a single EFT cannot describe parametrically large distances in moduli space, since a parametrically large number of massive particles inevitably become light, and we need UV information to determine their properties.

However, the conjecture places no restrictions on traversals that are larger than $M_{\rm Pl}$, but not parametrically large. For instance, there are examples in which the density of states is essentially constant along a trajectory of super-Planckian length.\footnote{Such a trajectory is not necessarily a geodesic in moduli space~\cite{Hebecker:2017lxm}.} This typically occurs when the modulus in question is an axion, consistent with Conjecture 2 because the compact field space prevents an arbitrarily large excursion. While axion excursions do not lead to a tower of particles becoming light, we still expect a tower of particles with masses that change as the axion expectation value varies. When the axion field completes a full circuit, the spectrum must return to where it started, perhaps with a nontrivial monodromy. 

Motivated by axions, we reconsider the effect of a tower of particles with masses $m_n(\phi)$ that depend on a modulus $\phi$, dropping the assumption that the masses vanish as $\phi \rightarrow 0$. By the same analysis as above, if the modulus becomes strongly coupled at the quantum gravity scale $\Lambda_{\rm QG}$, i.e., if $\lambda_{\phi}(\Lambda_{\rm QG}) \sim 1$, then the modulus kinetic term is approximately
\begin{align}
K(\phi) \sim \Lambda_{\rm QG}(\phi)^{D-4} \sum_{n \, | \, m_n(\phi) < p}
  \left( \frac{\partial m_n}{\partial \phi} \right)^2.
  \label{eq:axionK}
\end{align}
While in general both $K(\phi)$ and $\Lambda_{\rm QG}(\phi)$ will depend on the modulus, if $\phi$ is an axion then typically both are approximately constant---roughly speaking because the axion has an approximate shift symmetry. In this case, by averaging over all particles with mass below the cutoff, we obtain
\begin{align}
\Lambda_{\rm QG}^{D-4} N(\Lambda_{\rm QG})  \left< \left(\frac{\partial m_n}{\partial \phi}\right)^2 \right>  \sim K.
\end{align}
We can eliminate $N(\Lambda_{\rm QG})$ from this expression using the species bound (\ref{eq:speciesbound}). Under a traversal $\phi \rightarrow \phi+ \Delta \phi$, we find\footnote{Here we assume that the spectrum is not rapidly oscillating as a function of $\phi$ in order to approximate $\frac{\partial m_n}{\partial \phi}\sim \frac{\Delta m_n}{\Delta \phi}$.}
\begin{equation} \label{eq:axiontraversal}
   \langle (\Delta m_n)^2 \rangle  \sim   \frac{(\Delta \phi)^2 K}{M_{\rm Pl}^{D-2}} \Lambda_{\rm QG}^2.
\end{equation}
Since $\Delta \phi \sqrt{K}/M_{\rm Pl}^{(D-2)/2}$ is the length of the traversal in Planck units, we learn that as $\phi$ rolls a Planckian distance in field space, the typical massive mode shifts by an amount of order the quantum gravity scale $\Lambda_{\rm QG}$.

This is reminiscent of a phenomenon observed in \cite{Heidenreich:2015wga}: in certain models of large-field axion inflation (including some models of decay constant alignment \cite{kim:2004rp} and axion monodromy \cite{mcallister:2008hb,silverstein:2008sg}), modes that begin above the UV cutoff become very light as the inflaton rolls. However, \eqref{eq:axiontraversal} does not necessarily imply that modes above $\Lambda_{\rm QG}$ become very light under a super-Planckian traversal $\Delta \phi \sqrt{K} > M_{\rm Pl}^{(D-2)/2}$: a mode with mass of order $\Lambda_{\rm QG}$ generically acquires a different mass of order $\Lambda_{\rm QG}$. On the other hand, \eqref{eq:axiontraversal} does imply that generically an order-one fraction of the modes will pass through the cutoff during this traversal.

To illustrate these points, consider an axion $\theta$ arising from compactification of a $5$-dimensional $U(1)$ gauge theory with coupling constant $e_5$ on a circle of radius $R$. Consistent with recent ideas about the WGC~\cite{Heidenreich:2015nta,Heidenreich:2016aqi,Montero:2016tif,Andriolo:2018lvp} and the analysis of~\cite{Heidenreich:2017sim}, we assume the 5d theory has a tower of near-extremal charged particles, which leads to a 4d KK spectrum of the form
\begin{align}
m_{n_1,n_2}^2 \sim  \left( n_1^2 e_1^2 + e_2^2 (n_2 - n_1 \theta)^2  \right) M_{\rm Pl}^2.
\label{eq:axionmass}
\end{align}
Here $e_1^2 = e_5^2/(2 \pi R)$ and $e_2^2 = 2/(R M_{\rm Pl})^2$ are the 4d gauge couplings and $\theta \cong \theta +1$ is the axion. Assuming a sufficiently large number of modes charged under each $U(1)$ lie below the cutoff, we can approximate the sums over $n_1$ and $n_2$ in (\ref{eq:axionK}) as integrals, giving
\begin{align}
(2 \pi f)^2 \df K \sim (e_2/e_1)^2  M_{\rm Pl}^2,
\end{align}
in agreement with tree-level dimensional reduction, where we express the result in terms of the ``axion decay constant'' $f$.

Assuming a particular mode has mass $m_{n_1,n_2} \sim \Lambda_{\rm QG}$ when $\theta =0$, what is the lightest this mode can become under $\theta \rightarrow \theta + \Delta \theta$? The ideal situation occurs when $n_2 \approx n_1 \Delta \theta$, so that the second term in the mass formula (\ref{eq:axionmass}) becomes negligible after the shift. Setting $m_{n_1,n_2}^2 \sim e_2^2 n_2^2 M_{\rm Pl}^2 \sim \Lambda_{\rm QG}^2$ at $\theta=0$, we find a final mass of
\begin{align}
m_f^2 \sim e_1^2 n_1^2 M_{\rm Pl}^2 \sim  \left( \frac{M_{\rm Pl}}{2 \pi f \Delta \theta}\right)^2 \Lambda_{\rm QG}^2 .
\end{align}
Thus, the mass of the lightest mode that begins above the cutoff $\Lambda_{\rm QG}$ is inversely proportional to the axion shift in Planck units. For a modest super-Planckian traversal $2 \pi f \Delta \theta \sim \mbox{10--100}\ M_{\rm Pl}$ (as required for large-field axion inflation), this mass is not very light.

To estimate the number of modes passing through the cutoff as $\theta \to \theta+ \Delta \theta$, note that for fixed $n_1$, $2n_1$ modes pass upwards/downwards through the cutoff over a full period, $\Delta \theta = 1$. Thus, the total number of modes passing through the cutoff is $\Delta N \sim \Delta \theta \sum_{n_1} 2 n_1 \sim \Delta\theta (\Lambda_{\rm QG}/e_1 M_{\rm Pl})^2$. Comparing with the total number of light modes, we obtain
\begin{equation}
\frac{\Delta N}{N} \sim \frac{2 \pi f \Delta \theta}{M_{\rm Pl}}.
\end{equation}
Thus, for a Planckian field traversal, an order-one fraction of the modes will pass through the cutoff, and for a larger field traversal almost all the modes will be recycled.

What does this mean for EFT and axion inflation? On the one hand, we could integrate out all the modes with mass $e_1 M_{\rm Pl}$ or above, obtaining an EFT with a lower cutoff but with no apparent drama as the axion traverses a super-Planckian distance. On the other hand, if we wish to compute the axion potential using EFT then we cannot take this approach, since in this example the axion potential is generated by the Casimir energy of charged particles in the 5d parent theory~\cite{Hosotani:1983xw,Cheng:2002iz,ArkaniHamed:2003wu}, whose masses start at $e_1 M_{\rm Pl}$. If we raise the cutoff to include some of these charged particles in the EFT then we once again face the twin issues of modes emerging from the cutoff and becoming light and a large fraction of the modes passing through the cutoff during the axion traversal.\footnote{In the language of 5d EFT, these issues are related to the difficulty of imposing a gauge-invariant cutoff on loops of charged particles.} Both issues suggest that an EFT incorporating these modes is not well controlled for a super-Planckian field excursion in the absence of additional UV input.

Thus, there are potential subtleties in using an EFT (even a string theory-derived EFT) to compute the potential of an axion over a very super-Planckian field range. Our arguments suggest that these subtleties extend beyond the extra-natural context explored in~\cite{Heidenreich:2015wga} and to other moduli besides axions. It remains to be seen whether these subtleties are of critical importance in candidate large field models.

\section{Modulus Potential}

Above we assumed a light modulus that can be described within EFT. 
We should test this assumption: supersymmetric theories can have exactly massless moduli, but more generally, do loops tend to give moduli a large mass? If we sum up the loop corrections to a modulus from a tower of fermions, we find power-divergent diagrams. Cutting these off at $\Lambda_{\rm QG}$, we have
\begin{equation}
\delta m_{\rm mod}^2 \sim\!\!\!\! \sum_{n\, | \, m_n < \Lambda_{\rm QG}}\! \frac{m_n^2}{M_{\rm Pl}^{D-2}} \Lambda_{\rm QG}^{D-2} \sim \frac{N(\Lambda_{\rm QG}) \Lambda_{\rm QG}^D}{M_{\rm Pl}^{D-2}} \sim \Lambda_{\rm QG}^2.
\end{equation}
However, in the presence of supersymmetry, contributions from scalars and their fermionic partners approximately cancel to leave a remainder of order the average SUSY-breaking splitting within the tower of states:
\begin{equation}
\left.\delta m_{\rm mod}^2\right|_{\rm SUSY} \sim \left<m_{\rm boson}^2 - m_{\rm fermion}^2\right>.
\end{equation}
In the absence of supersymmetry, loop corrections will drive the modulus mass to the cutoff $\Lambda_{\rm QG}$. This must be reconciled with examples, like compactification of a nonsupersymmetric theory on a circle, in which we know that a radion modulus exists with a controlled (Casimir) potential. The resolution of this puzzle is that tuning away the cosmological constant in the parent theory in turn fine-tunes the modulus potential of the lower-dimensional theory; a c.c.~of the naive size in the parent theory becomes precisely a modulus potential with curvature $\Lambda_{\rm QG}^2$ in the daughter theory.

In supersymmetric theories, the naive expectation is that moduli masses are of order the gravitino mass. It is known that no-scale structure allows lighter moduli \cite{Cremmer:1983bf, Luty:1999cz, Balasubramanian:2005zx}. For consistency with the estimate above, this means that the tower of states with masses controlled by the no-scale modulus should {\em also} have small SUSY breaking. We can check this explicitly in theories where no-scale structure arises from the overall volume modulus of a compactification, which happens in 4d theories when reducing from 5d or 10d Type IIB supergravity \cite{Reece:2015qbf}. No-scale structure is manifest in the ``keinstein'' frame where the modulus chiral superfield $\bm{T}$ obtains a kinetic term entirely by mixing with the graviton, appearing as $\int d^4 \theta M_*^3 \bm{\Phi^\dagger \Phi} (\bm{T^\dagger} + \bm{T})$. Here $\bm{\Phi}$ is the conformal compensator chiral superfield and the linear coupling to $\bm{T}$ naturally produces $|F_\Phi / \Phi| \ll m_{3/2}$. One can show that in keinstein frame, a field $\bm{\chi}$ propagating in the bulk in the higher dimensional theory has KK modes in the 4d theory with K\"ahler potential $\int d^4 \theta \bm{\Phi^\dagger \Phi} (\bm{T^\dagger} + \bm{T}) \bm{\chi_n^\dagger \chi_n}$, giving rise to soft masses $\widetilde{m}_{\chi_n}^2 \sim |F_\Phi/\Phi| |F_T/T| \sim m_{3/2} |F_\Phi / \Phi| \ll m_{3/2}^2$. In other words, no-scale protection from large SUSY breaking naturally extends to the Kaluza-Klein modes of bulk fields. This shows that no-scale structure is  consistent with our simple estimate of summing loop corrections from the KK tower.

\section{Conclusions}

The Sublattice WGC \cite{Heidenreich:2015nta, Heidenreich:2016aqi, Montero:2016tif} and the recently proposed Tower WGC \cite{Andriolo:2018lvp}---strengthened versions of the Weak Gravity Conjecture motivated by dimensional reduction and evidence from string theory---require an infinite tower of light particles, closely linking them to the Swampland Conjectures. In some cases, the connection is direct: the gauge coupling $g$ is related to the vev of a scalar modulus and the $g \to 0$ limit brings down a single tower of light charged particles, satisfying both conjectures.

In other cases the Swampland Conjectures strengthen the Sublattice WGC by demanding that a tower of particles can be accessed within EFT. For instance, in \cite{Heidenreich:2017sim} we observed that in some examples the WGC tower is predicted to lie above $\Lambda_{\rm QG}$. A concrete case is an approximately isotropic 4d compactification of Type IIB string theory with gauge fields on D7 branes. The gauge coupling $g \sim 1/(M_s R)^2$, so the WGC tower is at $g M_{\rm Pl}$; the string scale is lower, at $g^{3/2} M_{\rm Pl}$; but the Kaluza-Klein tower is lower still, at $g^2 M_{\rm Pl}$. The WGC tower is outside the low-energy EFT, but the KK tower is not. It accounts for the infinite distance in moduli space as $g \to 0$ and the generation of strongly coupled gravity via the species bound. This phenomenon, with multiple towers of particles becoming light at different rates as one moves to infinity in moduli space, can arise in a variety of examples.

We have argued that the assumption of a universal strong coupling scale for fields in quantum gravity can serve as a more fundamental replacement for some of the Swampland Conjectures. It is still important to put the most basic aspects of these conjectures on a firmer footing: can we prove rigorously that moduli exist and that large-moduli limits always send infinite towers of particles to zero mass? These basic assumptions have a similar flavor to the statement that quantum gravity theories have no global symmetries, and deserve more attention.

\begin{acknowledgments}
\vspace*{1cm}
{\bf Acknowledgments.}  The research of BH was supported by Perimeter Institute for Theoretical Physics. Research at Perimeter Institute is supported by the Government of Canada through the Department of Innovation, Science and Economic Development, and by the Province of Ontario through the Ministry of Research, Innovation and Science. MR is supported in part by the DOE Grant {DE-SC}0013607 and the NASA ATP Grant NNX16AI12G. TR is supported by the Carl P. Feinberg Founders' Circle Membership and the NSF Grant PHY-1314311.
\end{acknowledgments}

\bibliography{refs}

\end{document}